\documentclass{PoS}

\usepackage{amsmath}
\usepackage{amssymb}
\usepackage{graphicx}
\usepackage{slashed}
\usepackage{amstext}
\usepackage{amsfonts}
\usepackage{amscd}
\usepackage{latexsym}
\usepackage{verbatim}
\usepackage{mathbbol}

\newcommand{\bfig}{\begin{figure}}
\newcommand{\efig}{\end{figure}}
\newcommand{\bcom}{}

\newcommand{\as}{\alpha_s}
\newcommand{\alphas}{\alpha_s}
\newcommand{\asPi}{\frac{\alpha_s}{\pi}}

\newcommand{\mcbar}{\bar{m}_c}
\newcommand{\mubar}{\bar{\mu}}

\newcommand{\cbar}{\bar{c}}

\newcommand{\mcr}{\ensuremath{m_{\text{cr}}}}

\newcommand{\brackets}[1]{\langle #1 \rangle}

\newcommand{\rbrackets}[1]{\left( #1 \right)}

\newcommand{\psib}{\bar{\psi}}

\newcommand{\chib}{\bar{\chi}}

\newcommand{\gammamu}{\gamma_{\mu}}

\newcommand{\gammafive}{\gamma_{5}}

\newcommand{\tauthree}{\tau^3}
\newcommand{\taupm}{\tau^\pm}

\newcommand{\fps}{f_{PS}}
\newcommand{\fpi}{f_{\pi}}
\newcommand{\mpi}{m_{\pi}}
\newcommand{\mps}{m_{PS}}
\newcommand{\mjpsi}{m_{J/\psi}}
\newcommand{\mdpm}{m_{D^{\pm}}}
\newcommand{\metac}{m_{\eta_c}}

\newcommand{\mfps}{\frac{\mps}{\fps}}

\newcommand{\bitem}{\begin{itemize}}
\newcommand{\eitem}{\end{itemize}}
\newcommand{\bnum}{\begin{enumerate}}
\newcommand{\enum}{\end{enumerate}}
\newcommand{\ba}{\begin{eqnarray}}
\newcommand{\ea}{\end{eqnarray}}
\newcommand{\bcen}{\begin{center}}
\newcommand{\ecen}{\end{center}}

\newcommand{\fermi}{\,\text{fm}}
\newcommand{\gev}{\,\text{GeV}}
\newcommand{\mev}{\,\text{MeV}}

\newcommand{\order}[1]{\mathcal{O}\left( #1 \right)}
\newcommand{\pvec}{\vec{p}}

\newcommand{\MSbar}{\overline{\text{MS}}}

\newcommand{\flavorone}{\mathbb{1}}

\newcommand{\epow}[1]{\text{e}^{#1}}

\newcommand{\localpath}{./}

\newcommand{\figurepath}{\localpath/figures}

\FullConference{The XXIX International Symposium on Lattice Field Theory - Lattice 2011\\
  July 10-16, 2011\\
  Squaw Valley, Lake Tahoe, California}
\title{\normalsize Charm Current-Current Correlators in Twisted Mass Lattice QCD}
\ShortTitle{Current correlators in tmLQCD}

\author{Karl Jansen\\
  NIC / DESY Zeuthen\footnote{Preprint number: DESY 11-198\\ SFB-Number: SFB/CPP-11-60} \\
  Platanenallee 6 \\
  D-15738 Zeuthen \\
  Germany \\
  E-mail: \email{Karl.Jansen@desy.de}}

\author{\speaker{Marcus Petschlies}\\
  Humboldt-Universit\"at zu Berlin, Institut f\"ur Physik \footnote{Preprint number: HU-EP-11/49}\\
  Newtonstra{\ss}e 15\\
  12489 Berlin \\
  E-mail: \email{marcuspe@physik.hu-berlin.de}}

\author{Carsten Urbach\\
  HISKP (Theory) and  Bethe Center for Theoretical Physics \\
  Rheinische Friedrich-Wilhelms-Universit{\"a}t Bonn \\
  Nussalle 14-16, 53115 Bonn, Germany\\
  E-mail: \email{urbach@hiskp.uni-bonn.de}}

\author{for the European Twisted Mass collaboration}

\abstract{
  The current correlator method has been shown to be a practical tool
  to extract the charm quark mass and strong coupling constant  
  from Lattice QCD data as an alternative to the sum rule approach
  using experimental electron-positron annihilation cross section
  data. We report on the progress of an investigation of charm
  current-current correlators in $N_f=2$ Twisted Mass Lattice QCD. Upon
  determining the temporal moments of the current correlators we
  compare to the low-energy expansion of the moments in perturbative
  QCD and calculate the charm quark mass and strong coupling constant
  and in case of the vector current correlator directly compare both
  methods.
}

\FullConference{The XXIX International Symposium on Lattice Field Theory, Lattice2011\\
  July 11-16, 2011\\
  The Village at Squaw Valley, USA}

\begin{document}

\section{Introduction}
\label{sec:intro}

The charm quark mass and strong coupling constant are fundamental
parameters of the Standard Model and thus there is an interest per se
in their calculation. They are essential input parameters for the
calculation of processes involving charm quarks, such as inclusive
radiative $B$-decays and exclusive Kaon-decays
\cite{Dehnadi:2011gc}. Moreover they play an important role in the
estimation of CKM matrix elements and the search for new physics
beyond the Standard Model \cite{Antonelli:2009ws}.

Recently the HPQCD collaboration extracted the $\MSbar$ charm quark
mass and strong coupling constant using temporal moments of charmed
lattice current correlators~\cite{McNeile:2010ji}. Using the highly
improved staggered 
quark action and a Bayesian prior fitting analysis a few percent
precision could be reached. Here we report on an ongoing effort to
apply this method using a different fermion discretization, namely
Wilson twisted mass Lattice QCD.
In this work, we will not rely on any Bayesian prior in the fits of our data
and it is one of our goals to understand, whether a similar accuracy
can be reached as given in \cite{McNeile:2010ji}.

\section{Low momentum expansion of polarization functions in perturbative QCD}
\label{sec:low_momentum_pQCD}

The general strategy of the current correlator method is the
non-perturbative estimation of derivatives $M_n$ of the polarization
functions of in our case the  pseudoscalar and vector currents from
lattice data and to compare them to their continuum counterparts
determined in perturbation theory. The derivatives are readily deduced
from the momentum expansion of the polarization functions in the limit
$q^2 \ll \mcbar^2$
\begin{eqnarray}
  q^2\,\Pi_c^{\kappa} &=& i \int d^4x \epow{iqx} \brackets{ 0 |
    T\{J^{\kappa} (x)\,J^{\kappa} (0)\}\,|0 }\,, \nonumber\\ 
  \left(-q^2 g_{\mu\nu} + q_\mu q_\nu \right) \Pi_c^{\delta} + q_\mu
  q_\nu \Pi_{c\,L}^{\delta} &=& i \int d^4x\, \epow{iqx}  
  \brackets{ 0 | T\{J_\mu^{\delta} (x)\,J_\nu^{\delta} (0)\}|0 }\,,
  \label{eq:continuum_vacuum_polarization}\\ 
  \Pi^{\kappa,\delta}(q^2) &=& \frac{3}{16\pi^2}\sum\limits_{k\ge -1}
  \bar{C}^{\kappa,\delta}_k\, z^k\,, \nonumber \\ 
  \bar{C}^{\kappa,\delta}_k  &=& \sum\limits_{m\ge 0} \,\left(
    \frac{\alpha_s}{\pi}\right)^m \,
  \bar{C}^{\kappa,\delta\,(m)}_k \, \left(\log\left( \frac{\mcbar^2(\mu)}{\mu^2} \right)
  \right)\,, \label{eq:pQCD_low_momentum_expansion} \\
  M^{\kappa,\delta}_n &=& \left. \frac{12\pi^2}{n!}\rbrackets{\frac{d}{dq^2}}^n \Pi^{\kappa,\delta}_c(q^2) \right|_{q^2=0}
  \label{eq:pQCD_moment_definition}
\end{eqnarray}
with $\delta = v,a$, $\kappa= p,s$, $J^{p} = \cbar \gamma_5 c$, $J^{s}
= \cbar c$, $J_{\mu}^{v} = \cbar  \gammamu c$, $J_{\mu}^{a} = \cbar
\gammamu \gamma_5 c$. The perturbative expansion of the coefficients
$\bar{C}^{\kappa,\delta}_k$ has nowadays reached the 4 loop level
($\order{\as^3}$) ( cf. \cite{Kiyo:2009gb} and references therein).

\section{Lattice Formulation}
\label{sec:setup}

The calculation we report on here is based on gauge configurations
produced by the European Twisted Mass collaboration (ETMC) using
$N_f=2$ flavors of maximally twisted and mass degenerate Wilson fermions. We
refer the reader to ref.~\cite{Baron:2009wt} and references therein.
We treat the charm degrees of freedom in a partially quenched
framework by adding a doublet of heavy quarks $\chi =
(\chi_+,\,\chi_-)$ in the valence sector with valence quark action
\begin{equation}
  \label{eq:tm_heavy_quark_action}
  \mathcal{S}_{val} = \sum\limits_x \chib(x) \left( D_W + \mcr +
    i\mu_h \gammafive\tauthree \right)\chi(x)\,. 
\end{equation}
In this framework automatic $\order{a}$
improvement~\cite{Frezzotti:2003ni} is in place with the same critical
Wilson mass $\mcr$ as used in the light quark sector. For the lattice 
operators representing the physical charm currents for a given spin
structure $\Gamma$ we have three natural choices (given in the
physical basis). 
\begin{equation}
  \label{eq:lattice_charm_currents}
  J^0_\Gamma = \psib\,\Gamma\otimes\flavorone\,\psi\,; \quad
  J^3_\Gamma = \psib\,\Gamma\otimes\tauthree\,\psi\,; \quad 
  J^\pm_\Gamma = \psib\,\Gamma\otimes\taupm\,\psi\,.
\end{equation}
At non-zero lattice spacing the correlation of these operators with
themselves will give a different results for each operator due to
lattice artifacts. Concerning the physical charm fields we would need
to use the singlet currents and their corresponding translation in
terms of the $\chi$ fields in the twisted basis. However, in our
calculation we will not consider contributions from quark-disconnected
diagrams. This is not a source of error given the fact that the
perturbative expressions we will compare to will not include
singlet contributions as well (entering at $\order{\as^3}$ for the
vector and $\order{\as^2}$ for the pseudoscalar currents). But given
the absence of quark-disconnected diagrams the two-point correlator of
$J^0_\Gamma$ will coincide with that of $J^3_\Gamma$. In the continuum
limit vector flavor symmetry restoration will entail the latter to
become equal to the correlation function of $J^\pm_\Gamma$. This
circumstance allows us to exploit the features of tmLQCD when it comes
to the multiplicative renormalization of the bare current correlators
to our advantage.

In terms of the currents defined above the renormalized and
dimensionless vector and pseudoscalar moments read in the twisted
basis 
\begin{eqnarray}
  G^{V}_{n} &=&  Z_V^2 a^6\sum\limits_{t/a=-N_t/2+1}^{N_t/2-1} \left(t/a\right)^n \brackets{J_V^{0/3}\,J_V^{0/3} (t,\pvec=0)}^{conn} \nonumber \\
  &=&  Z_A^2 a^6\sum\limits_{t/a=-N_t/2+1}^{N_t/2-1} \left(t/a\right)^n \brackets{J_A^{\pm}\,J_A^{\mp} (t,\pvec=0)} \label{eq:vector_moment_def} \\
  G^{P}_{n} &=&  (2a\mu_h)^2\rbrackets{\frac{Z_S}{Z_P}}^2 a^6\sum\limits_{t/a=-N_t/2+1}^{N_t/2-1} \left(t/a\right)^n
  \brackets{J_S^{0/3}\,J_S^{0/3} (t,\pvec=0)}^{conn} \nonumber \\
  &=&  (2a\mu_h)^2 a^6\sum\limits_{t/a=-N_t/2+1}^{N_t/2-1} \left(t/a\right)^n
  \brackets{J_P^{\pm}\,J_P^{\mp} (t,\pvec=0)}\,. \label{eq:pscalar_moment_def}
\end{eqnarray}
Using $\mu_h^R = \mu_h/Z_P$ we introduced additional factors of
$a\mu_h$ such that only the scale independent ratio $Z_S/Z_P$ is
needed for the scalar moments and no renormalization factor for the
pseudoscalar moments. For the scale independent renormalization factors
$Z_P/Z_S,\,Z_A,\,Z_V$ we use the non-perturbative renormalization data
provided by ETMC (\cite{Constantinou:2010gr} and private communication).

The ensembles we choose comprise four different
lattice spacings ranging from $a \approx 0.05 \fermi$ to $a \approx
0.1 \fermi$ and light pseudoscalar masses in the range $280 \mev
\lesssim \mps \lesssim 650 \mev$ as well as up to two lattice volumes.
For each triple $(a,\,L,\,\mps)$ the current two-point functions were
measured with four to seven charm valence quark masses such that the
charmed meson masses $a\mjpsi/a\fps$, $a\metac/a\fps$  and
$a\mdpm/a\fps$ in units of the light pseudoscalar decay constant
covered the physical value \cite{Nakamura:2010zzi}. We
are thus able to study the dependence of the moments $G =
G(a,L,\mps,\mu_c)$ on all lattice parameters. The values of the light
pseudoscalar decay constants at the physical point for all four
lattice spacings were calculated in a separate dedicated fit 
along the lines of \cite{Baron:2009wt}.

\section{Analysis and results}
\label{sec:analysis}

\subsection{General outline}
\label{subsec:outline}

In our analysis we will model the dependence of the moments on the
lattice parameters $\mu_q$, $\mu_c$ and $a$. For extrapolating to the
physical light quark mass we shall use the charged pion mass $\mps$,
for interpolating to the physical charm quark mass the ground
state mass determined from the $c\cbar$ non-singlet vector current
correlator $\mjpsi$ and the lattice spacing dependence will be studied
using $a/\fps$. Finite volume effects turn out to be negligible in the charm sector.

We shall use two methods:
\begin{itemize}
\item interpolate the lattice data at each value of the lattice
  spacing to common reference points
  $\rbrackets{\rbrackets{\mps/\fps},
    \rbrackets{\mjpsi/\fps}}^{\mathrm{ref}}$. This strategy we shall
  denote with "$ref$".

\item perform a combined fit to our data describing the combined
  $(\mps/\fps,\,\mjpsi/\fps,\, a)$ dependence. This method we shall denote with
  "$all$" and it is based on splitting the fit function into a
  continuum part and one that models lattice artifacts as follows
  \begin{equation}
    \begin{split}
      \mathcal{F}\rbrackets{ a\fps,\, a\mps,\, a\mjpsi }& = \mathcal{F}_{cont} \times \mathcal{F}_{latt}
      = \sum\limits_{i=0}^{M}\sum\limits_{j=0}^{N} (a\mps/a\fps)^{2i} (a\mjpsi/a\fps)^j \\
      & \quad \times \rbrackets{ 1 + \sum_{\substack{0\le l,m,n\le 2\\ 0<l+m+n\le 2,\,4}}
        (a\fps)^{2l}(a\mps)^{2m}(a\mjpsi)^{2n} }\,.\\
      \label{eq:all_method_fit_function}
    \end{split}
  \end{equation}
  We then read of the value of the moments at zero lattice spacing and
  at the physical point by setting 
  \begin{equation}
    \label{eq:all_method_physpoint}
    \left. G \right|_{physical} = \mathcal{F}_{cont}\left(\mpi/\fpi,\, \mjpsi/\fpi\right) \,.
  \end{equation}
\end{itemize}
Either way we will end up with estimates for the continuum values of
the moments or functions thereof. With these estimates we can then set
up determining equations for the $\MSbar$ quark mass $\mcbar$ and the
strong coupling $\alphas$ using the perturbative representation of the
moments from the low momentum expansion of the polarization
functions. We thus set 
\begin{equation}
  \label{eq:moment_lqcd_pqcd_comp}
  \left. G \right|_{physical} = \sum\limits_{l=0}^{L}
  \rbrackets{\asPi}^{l} \bar{C}^{(l)}\rbrackets{\mcbar, \mubar} 
\end{equation}
where similar to $G$ the coefficients $C^{(l)}$ will be functions of
the original expansion coefficients in equation
\ref{eq:pQCD_low_momentum_expansion}. In the two cases we consider we
either use the charm quark mass or the strong coupling as input and
solve the equation for the remaining quantity. Errors are estimated
using a bootstrap method.


\subsection{Moments from the vector current correlator}
\label{subsec:vector_moments}

The moments of the vector current correlator provide a benchmark of
the method because their values are accessible using a dispersion
integral and measurements of the hadronic cross section ratio $R(s)$.
We can thus make a comparison of lattice and continuum data already on
the level of individual moments. The values of the continuum moments
we compare to have been provided by the authors of \cite{Kuhn:2007vp}
and recently in \cite{Dehnadi:2011gc} (cf. the detailed description of
the extraction of the charm piece in \cite{Kuhn:2007vp}).

\begin{figure}[t]
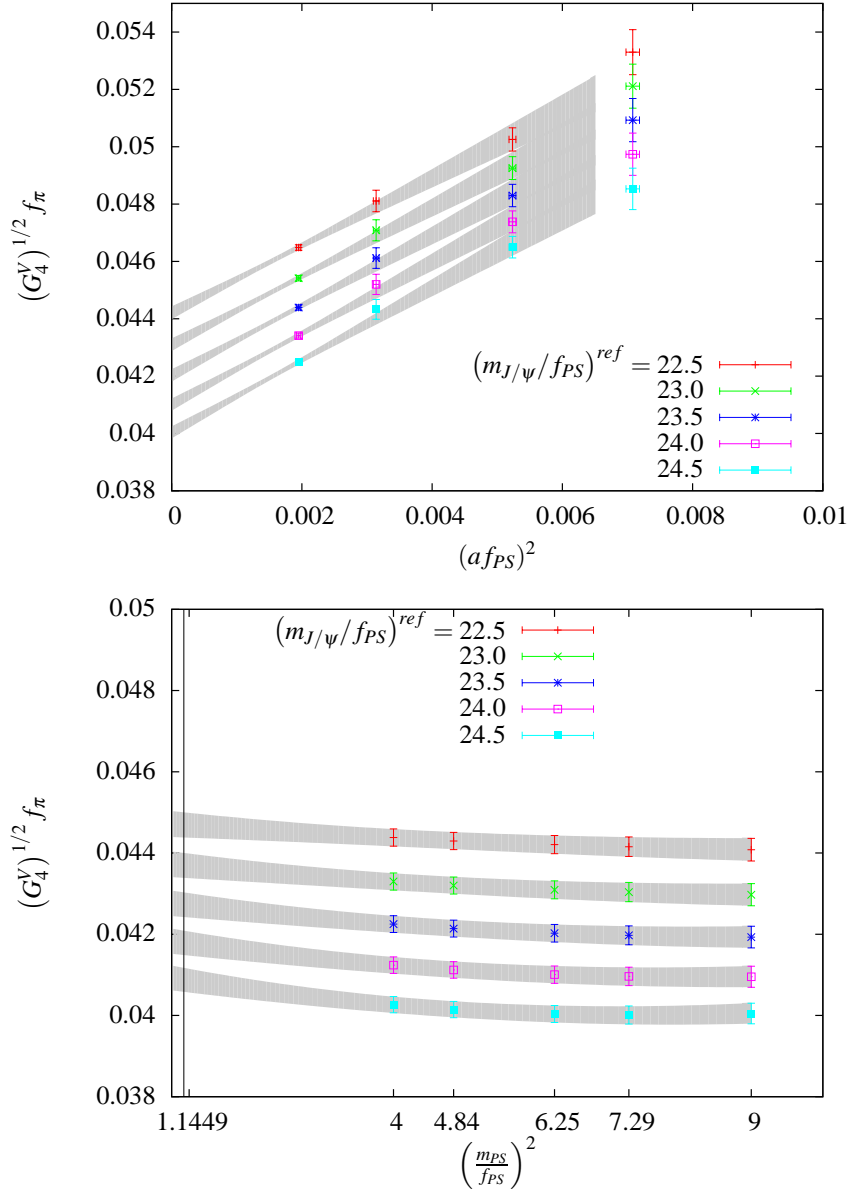

  \centering
  \graphicspath{{\figurepath/}}
  \scalebox{0.9}{\input{./figures/moments_5_s0_t00_m04_02_ll2.tex}}
  \scalebox{0.9}{\input{./figures/momcont_5s0_t00_m04_c02_res.tex}}
  \caption{Example for continuum and $\mps$ extrapolation of $G^V_4$
    with the $ref$ method.} 
  \label{fig:gv4_contexp_mpsexp}
\end{figure}

Taking into account all the explicit factors of the lattice spacing in
equation \ref{eq:vector_moment_def} dimensional analysis implies the
relation of dimensionless lattice moments $G^V_n$ and the
corresponding continuum quantities $g^V_n$ at non-zero lattice spacing
\begin{equation}
  \label{eq:vector_moments_lat_cont_artifacts}
  G^V_n = \frac{g^V_n}{(a\mcbar)^{n-2}} + \quad
  \mathrm{lattice~artifacts}\, .
\end{equation}
In figure \ref{fig:gv4_contexp_mpsexp} we show exemplary data for the
$ref$ method: the left panel shows the continuum extrapolation of the
vector moment $G^V_4$ at light pseudoscalar reference mass $a\mps /
a\fps = 2.5$ for five charm meson reference masses $a\mjpsi / a\fps =
22.5,\,23.0,\ldots,\,24.5$ (physical point at $\mjpsi / \fpi =
23.69\,(7)$ \cite{Nakamura:2010zzi}). The reference points with lower
light pseudoscalar masses than shown in the plot
($a\mps/a\fps=2.0,\,2.2$) are not entirely covered by the data from
the coarsest lattice which is why we leave it out of the extrapolation
and use a linear ansatz in $a^2$. The right-hand side panel shows the
extrapolation to the physical value of the light pseudoscalar mass
$\rbrackets{a\mps/a\fps}^{\mathrm{ref}} \to \mpi/\fpi = 1.068(3)$
\cite{Nakamura:2010zzi}. For the second extrapolation we again use a
polynomial ansatz of maximally second degree. The "$all$" method gives
comparable results.

\begin{table}[t]
  \centering
  \begin{tabular}{cccc}
    No. & $\fpi\left[ M^v_{n}(2n+2)!/(12\pi^2)/Q_c^2 \right]^{1/(2n)}$ &
    $G_{2n+2}^{V\,1/(2n)}f_\pi$ (ref) & $G_{2n+2}^{V\,1/(2n)}f_\pi$
    (all) \\ 
    \hline
    1 & 0.04107\,(32) & 0.04170\,(25) & 0.04146\,(77) \\
    2 & 0.08792\,(48) & 0.08810\,(52) & 0.08696\,(87) \\
    3 & 0.13081\,(60) & 0.13059\,(68) & 0.12945\,(94) \\
    4 & 0.17106\,(70) & 0.17098\,(82) & 0.16959\,(102)\\
    \hline
    \hline
  \end{tabular}
  \caption{Comparison of continuum vector moments with results from $ref$ and $all$ methods.}
  \label{tab:moments}
\end{table}

In table~\ref{tab:moments} we compare values our continuum
extrapolated results for the four lowest lattice moments at the
physical point to the continuum moments~\cite{Kuhn:2007vp} (second
column) determined using experimental data. Apart from the lowest
moment $M^v_1$ / $G^V_4$ we find good agreement between both the two
methods and the lattice and continuum moments.

By comparing to perturbation theory we are now able to extract the
$\MSbar$ charm quark mass. To that end we use the strong coupling as
an input parameter: 
starting from the PDG value $\alphas(\mu=M_Z,N_f=5) = 0.1184\,(7)$
\cite{Nakamura:2010zzi} we evolve it to $\alphas(\mu=3\gev,N_f=4) =
0.255(4)$ using the \textit{RunDec} program
\cite{Chetyrkin:2000yt}. The results for the solution for the four
lowest vector moments are collected in table
\ref{tab:msbar_charm_mass}. The first contribution to the uncertainty
stems from the statistical error of the moment extrapolation, the physical
scale ($\fpi$) and the value of $\alphas$. The second one represents
the systematic uncertainty from the choice of the renormalization
scale: it is obtained by matching lattice and continuum moments  at
$\mu=(3\pm1) \gev$ and evolving the result back to the reference scale
$\mu=3\gev$ using 4-loop evolution \cite{Kuhn:2007vp}.

If for each individual method we combine the quark masses from the
different moments (taking into account their strong correlation) we
find for the combined values 
\begin{equation}
  \mcbar(\mu=3\gev,N_f=3+1) = \left\{
    \begin{array}{lll}
      &0.979\,(09)\gev \: &(\mathrm{ref}) \\ &0.998\,(14)\gev \: &(\mathrm{all}) \\
    \end{array}
  \right.\,.
  \label{eq:msbar_charm_mass_combined}
\end{equation}
The results from both extrapolation methods turn out to be compatible with
the findings of reference \cite{McNeile:2010ji}, $\mcbar(\mu=3\gev,N_f=3+1) = 0.986\,(6)\gev$.

A consistency check with the lowest pseudoscalar moment using the
determined charm quark mass as input leads to a value of the strong
coupling in good agreement with the value used as input for the charm
mass. This will be discussed in detail elsewhere.

\section{Summary and Outlook}
\label{sec:outlook}

\begin{table}[t]
  \centering
  \begin{tabular}{ccc}
    No. & $\mcbar(\mu=3\gev) [\gev]$ (ref) & $\mcbar(\mu=3\gev)
    [\gev]$ (all) \\
    \hline
    1 & 0.971\,(09)\,(01) & 0.979\,(24)\,(01) \\
    2 & 0.981\,(10)\,(02) & 0.998\,(15)\,(02) \\
    3 & 0.990\,(10)\,(11) & 1.001\,(12)\,(11) \\
    4 & 1.014\,(08)\,(35) & 1.024\,(09)\,(34) \\
    \hline
    \hline
  \end{tabular}
  \caption{Comparison of results for the charm quark mass using $ref$
    and $all$ extrapolated vector moments.}
  \label{tab:msbar_charm_mass}
\end{table}

With this intermediate report we showed that within the twisted mass
formalism and with presently available statistics we can determine the
moments of the charm vector current correlator in agreement with
experimental results and with comparable uncertainty. Following two
different analysis methods we can extract the $\MSbar$ charm quark
mass from both methods and find agreement taking into account both the
statistical and systematic uncertainties. Yet from the comparison of
the central values of both analysis methods we infer that with the
presently available quality of data a systematic error of
$\mathcal{O}(20)\mev$ must be taken into account for the charm quark
mass value.

A consistency check with the lowest pseudoscalar moment using the
determined charm quark mass as input leads to a value of the strong
coupling in good agreement with the value used as input for the charm
mass. Currently we also investigate other methods to extract the
strong coupling from flavor singlet current diagrams as recently
presented in ref.~\cite{Chetyrkin:2011aa}.

As a next step it will be very interesting to apply the methods
discussed here to the $N_f=2+1+1$ gauge configurations of
ETMC~\cite{Baron:2010bv}.

We thank all members of ETMC for the most enjoyable
collaboration. This work is funded in part by the DFG within
SFB/TR9-03. The computing time was made available to us by
FZ-J{\"u}lich on JUROPA and JUGENE.

\bibliographystyle{h-physrev5}
\bibliography{bibliography}

\end{document}